# Dense III-V/Si Phase-Shifter Based Optical Phased Array

*Weiqiang Xie*, Tin Komljenovic, Jinxi Huang, Michael L. Davenport and John E. Bowers*

E-mail: weiqiangxie@ucsb.edu

Department of Electrical and Computer Engineering, University of California Santa Barbara, Santa Barbara, CA 93106, USA

**Abstract:** High-performance integrated optical phased arrays (OPAs) are crucial for building up energy-efficient solid-state beam steering systems on a chip. This work demonstrates a first heterogeneous III-V/Si phase-shifter based OPA that utilizes III-V multiple quantum well (MQW) structures to provide superior performance. Demonstrated phase shifters have $V_\pi$ of only 0.45 V and residual amplitude modulation (RAM) of only 0.15 dB at 1550 nm for 2π phase shift with low power consumption of less than 3 nW, which are, to the best our knowledge, the state-of-the-art performance with a record low operating voltage, low power consumption, and low RAM in on-chip OPAs. Their bandwidth is >1 GHz across 200 nm wavelength range. Wavelength tuning is used to demonstrate two-dimensional beam steering with a field of view of 28°×51° in the 32-channel OPA with an emitter-array pitch of 2 μm. Side-mode suppression ratio of 16 dB is also achieved. The high-performance heterogeneous OPA paves the way for scalable and efficient on-chip beam steerers with large steering range and high beam quality.

## 1. Introduction

Integrated optical phased arrays (OPAs) are essential for a number of applications such as light detection and ranging (LiDAR), free-space communications, and imaging and sensing [1-11]. The silicon photonics platform using mature complementary metal-oxide-semiconductor (CMOS) manufacturing is very attractive for making OPA-based systems for practical applications due to high volume capability and low cost. In on-chip OPAs, it is essential to integrate high-performance phase shifters with low drive voltage and low power consumption, low optical loss, and high bandwidth. More importantly, dense integration in OPAs is required to alleviate and overcome grating lobes and to maximize the effective aperture size towards a diffraction-limited beam.

A variety of silicon photonics OPAs have been demonstrated, using Si thermo-optic (TO) [3, 4, 7, 9, 10] or Si electro-optic (EO) phase shifters [5, 8, 12]. Ideally TO phase shifters have negligible optical loss, but have limited bandwidth, typically a few kHz, which impacts beam steering speed and consequently final device performance. In addition, high-power consumption and thermal crosstalk limit the integration density. On the other hand, reverse-biased PN-junction EO-type Si phase shifters are capable of high-speed operation (tens of GHz), low power consumption (<2 μW) [11], and dense integration. However, such phase shifters suffer from high RAM associated with 2π phase shift [11]. The limitation is inherent to the modulation mechanism that exploits plasma dispersion effect to change real and imaginary parts of the refractive index, which are coupled via Kramers–Kronig relations [12]. Moreover, the modulation efficiency in Si EO modulators is generally low (i.e. a high voltage–length product $V_{\pi}L$ typically 1-3 V·cm), resulting in high operation voltage or excessive lengths for a 2π phase shift [13]. Apart from Si-based TO an EO phase shifters, an alternative choice is to implement a heterogeneous III-V/Si for OPA phase shifters to improve device performance, since the III-V can leverage additional modulation effects, Franz-Keldysh or quantum-confined Stark effect, band-filling, and Pockels effects among others. In fact, such heterogeneous optical phase shifters have already been investigated in Mach–Zehnder interferometer (MZI) based modulators [14-19], and demonstrated low voltage operation ($V_{\pi}L$ ~ 0.05-0.2 V·cm) and a low optical loss (<1 dB for a 2π phase shift) [19]. Moreover, the heterogeneous III-V/Si integration approach could provide a promising solution for complete OPA-based systems such as LiDAR by integrating laser sources, amplifiers, OPAs, and detectors on a single chip. The largest remaining challenge is demonstrating such phase shifters in dense configurations necessary for state-of-the-art OPAs. This work, to the best of our knowledge, is the first such demonstration in which the pitch between neighboring phase shifters is only 4 μm, allowing for efficient and compact OPAs.

## 2. Design and Fabrication

In our demonstration, we incorporate a III-V material bonded on top of Si waveguides to form a heterogeneous III-V/Si phase-shifter array. A schematic design of the OPA is shown in **Figure 1**a. It comprises a $1\times N$ ($N$=32 in our test device) star coupler [20], a III-V/Si phase-shifter array with a length of 5 mm, and a 10-mm-long waveguide-grating antenna array. Note that all components in our OPA are designed for fundamental transverse-electric (TE) mode. The passive SOI contains a 1.0 μm buried oxide and a 500 nm Si layer on top, which is etched by 231 nm to form ridge waveguides. The star coupler is designed to couple ~95% of total power into waveguide array in a Gaussian distribution along the array [20], which also results in a nearly Gaussian beam profile in $\psi$ axis suppressing the close-in side-lobes. The phase-shifter array has a pitch of 4 μm and the details of the design can be found in Supplementary Information (see Section S1). For the grating emitter, we adopt a surface waveguide grating with a very shallow etch depth for a low emission strength so as to realize mm to cm scale of effective emission length. See Supplementary Information (Section S2) for the detailed design and characterization of the individual grating emitter. Note that the pitch of grating array, determining the maximum steering angle range before sidelobes interfere with the main lobe, can be different from the pitch of phase shifter array and we have both 4 μm and 2 μm grating pitch designs. The devices were fabricated on 100 mm SOI wafer using 248 nm DUV lithography and a III-V/Si heterogeneous integration technique [21, 22] (the fabrication details in Supplementary Information, see Section S3). Figure 1c shows a fabricated OPA with 32 channels.

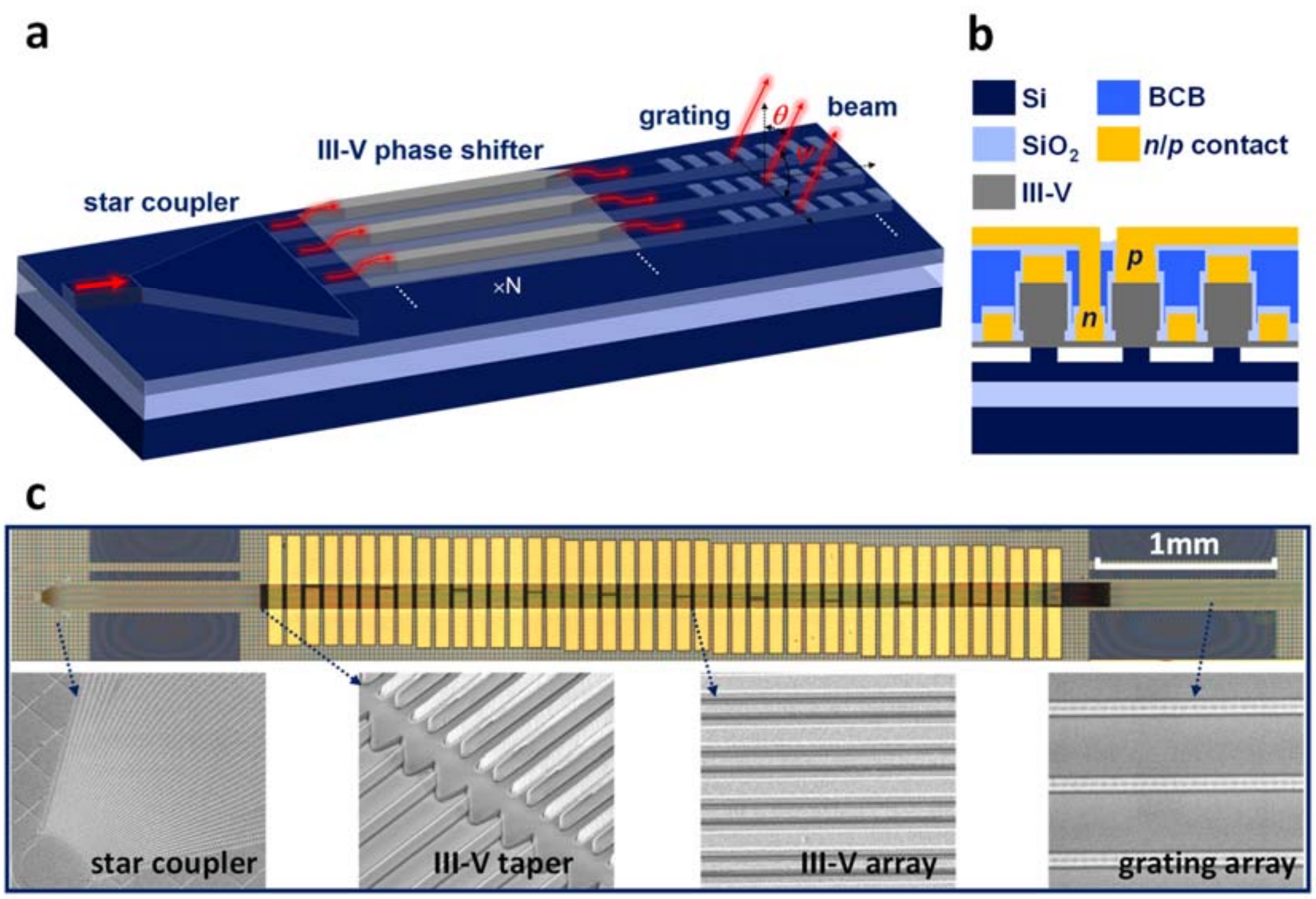

**Figure 1.** Design configuration and device fabrication. a**)** Schematics of a III-V/Si OPA system consisting of a 1×*N* star coupler, III-V/Si phase-shifter array, and waveguide-grating array. The input light is split into the Si waveguide array by a star coupler and then evanescently coupled to III-V/Si hybrid waveguide through an optimized low-loss III-V/Si taper. After the phase shift, the light is coupled back into Si waveguide layer and radiated coherently through the grating antenna array. The emitted beam direction in free space is denoted by $\theta$ and $\psi$, which are defined as the emission angles in waveguide dimension and OPA dimension respectively. b) Schematic cross-section of the III-V/Si phase shifters with locally opened *n*/*p* contact for probing in one shifter on top. c) Optical microscope image of a fully fabricated 32-channel OPA (upper panel) and SEM images of the corresponding components of the OPA (lower panel) taken before BCB planarization step.

## 3. Experimental Results

### 3.1. Characterization of III-V/Si Phase Shifter

The III-V/Si phase shifter was characterized using an unbalanced MZI with one arm having a 5mm heterogeneous phase shifter. The MZI comprises two 2×2 MMI couplers with 4 ports functioning as optical input and output respectively. By applying the voltage on III-V diode, we measured the transmission between one input port and one output port. **Figure 2**a shows the transmission at 1550 nm when sweeping the bias voltage. We obtained a reverse bias voltage of 0.9 V for a 2π shift and an extinction ratio larger than 23 dB. It can also be seen that the peak transmission intensity from 0 to -1V only reduces by ~0.1 dB, meaning a very low RAM within 2π phase shift. In Figure 2b, we present the bias voltage of a 2π phase shift as a function of wavelength and the operation voltage is in a range of 0.35 V to 1.4 V over a wavelength range of 1450 nm to 1650 nm. The absorption loss at 2π phase-shift voltage is also extracted by measuring the bias-dependent transmission of a III-V/Si waveguide and shown in Figure 2b. It can be seen that the loss penalty for the intensity is only ~0.15 dB for the wavelength from 1550 nm to 1650 nm, which implies a RAM of ~0.15dB. At shorter wavelengths, the loss increases due to band-edge absorption and can be alleviated by shifting the absorption edge of III-V QWs to even shorter wavelengths in future design. Considering the dark current of only 1-3 nA at -1 V bias (see Section S4, Supplementary Information), the power consumption of the phase shifter is on the order of nanowatts. The demonstrated state-of-the-art performance in the III-V/Si phase shifter with a record low operating voltage, low power consumption, and low RAM is suitable for on-chip OPAs. In addition, we measured the frequency response of the MZI modulator in Supplementary Information (see Section S4,) and obtain the 3dB electrical bandwidth of 1.65 GHz limited by capacitance. Such operation speed is sufficient for OPA applications.

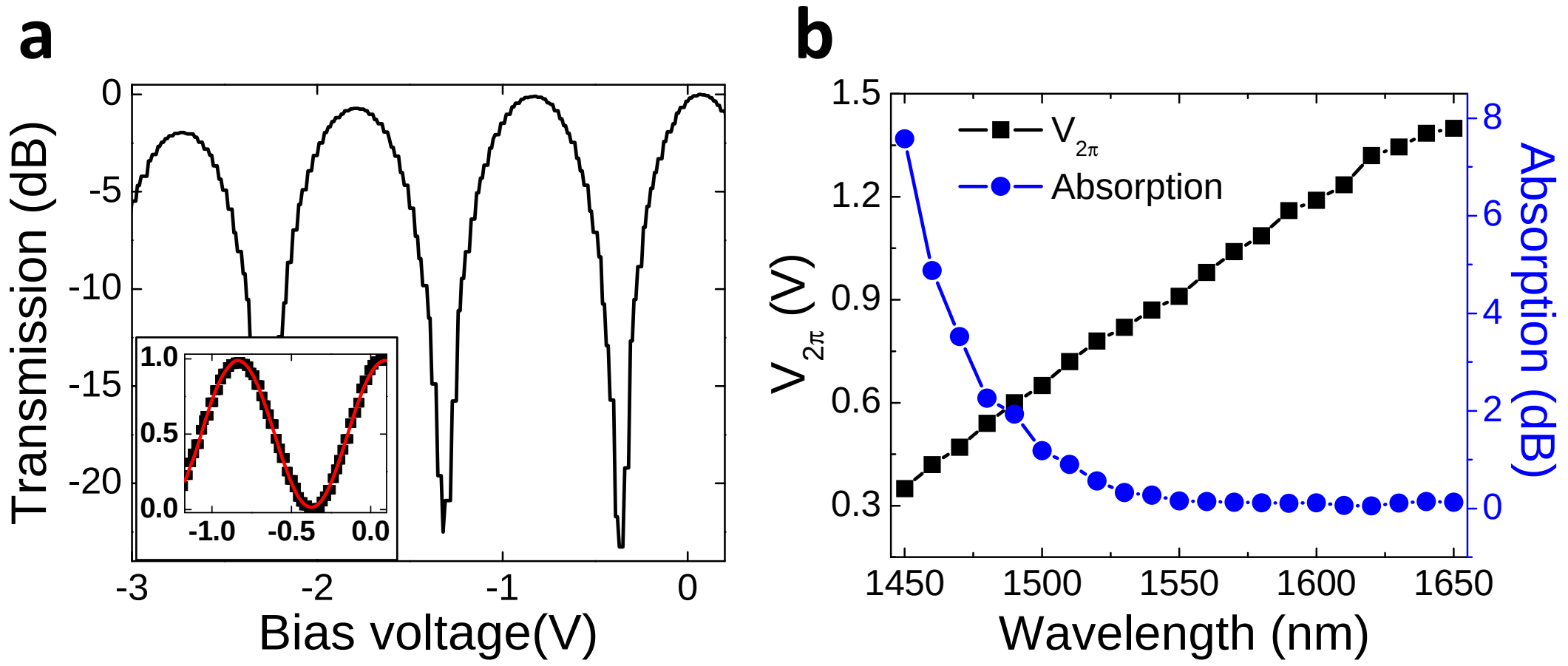

**Figure 2.** Static characterization of III-V/Si phase shifter. a) Transmission at 1550 nm between two MZI ports as a function of applied bias voltage on III-V diode. The inset shows the sine function fit of the transmission with a bias from 0 to -1 V, indicating a good linearity between applied bias and phase shift.

b) Reverse bias voltage (black) and total absorption loss (blue) at a 2π phase shift for different wavelengths in a 5mm phase shifter. Considering ~1 dB loss as maximum acceptable (which is till superior to Si phase shifters), the bandwidth of our III-V/Si phase shifter is ~150 nm from 1500 nm to 1650 nm.

### 3.2. Beam Steering Measurement of OPA

We first characterized the 32-channel OPA with a 4 μm pitch for the grating array using a Fourier imaging setup, as illustrated in **Figure 3**a. The phase shifters were reverse biased using 32 voltage sources controlled by a computer. The TE polarized light from a tunable laser was coupled into the star coupler and split into 32 waveguide channels, and an IR camera was used to acquire the far field image and to provide feedback for real-time phase optimization. The measured I-V curves as shown in Supplementary Information (see Section S4) reveals a uniform electrical performance among all phase shifters, which is important for the beam characterization. Without phase alignment, the beam emitted from the grating array is not focused due to path-length imbalance of the star coupler and accumulation of phase errors along the length of the device. To realize beam focusing and steering, we employed the gradient descent method to optimize the individual phases of the OPA. To implement the optimization, a target function is defined as $F = -P(\theta, \psi)/P_0$, where $P(\theta,\psi)$ is the beam power at desired $(\theta,\psi)$ direction and $P_0$ is the total power inside the first grating-lobe region. Figure 3b shows the evolution of the optimization function when steering the beam at $\psi = 0$. The optimization shows a very fast convergence; after only five iterations the voltages converge and the target function reaches 98% of the final optimized value. After phase alignment all bias voltages are in a range of 0 to -1 V, matching with the 0.9 V operation bias voltage for a 2π phase shift at 1550 nm. In Figure 3c, we plot the beam profiles along $\psi$ axis before and after phase optimization and obtain a side-mode suppression ratio of 16 dB for the focused beam, with the side-mode suppression ratio defined in a single grating-lobe region. It is also evident that for the 4 μm pitch grating array, the maximum beam steering angle in $\psi$ axis is ~22° in which the grating lobes are not present. The steering range is ~51° with a 2 μm pitch grating array at 1550 nm ($\sin\psi_{max} \approx \lambda/p_g$, with $\lambda$ the free-space wavelength and $p_g$ the pitch of grating array).

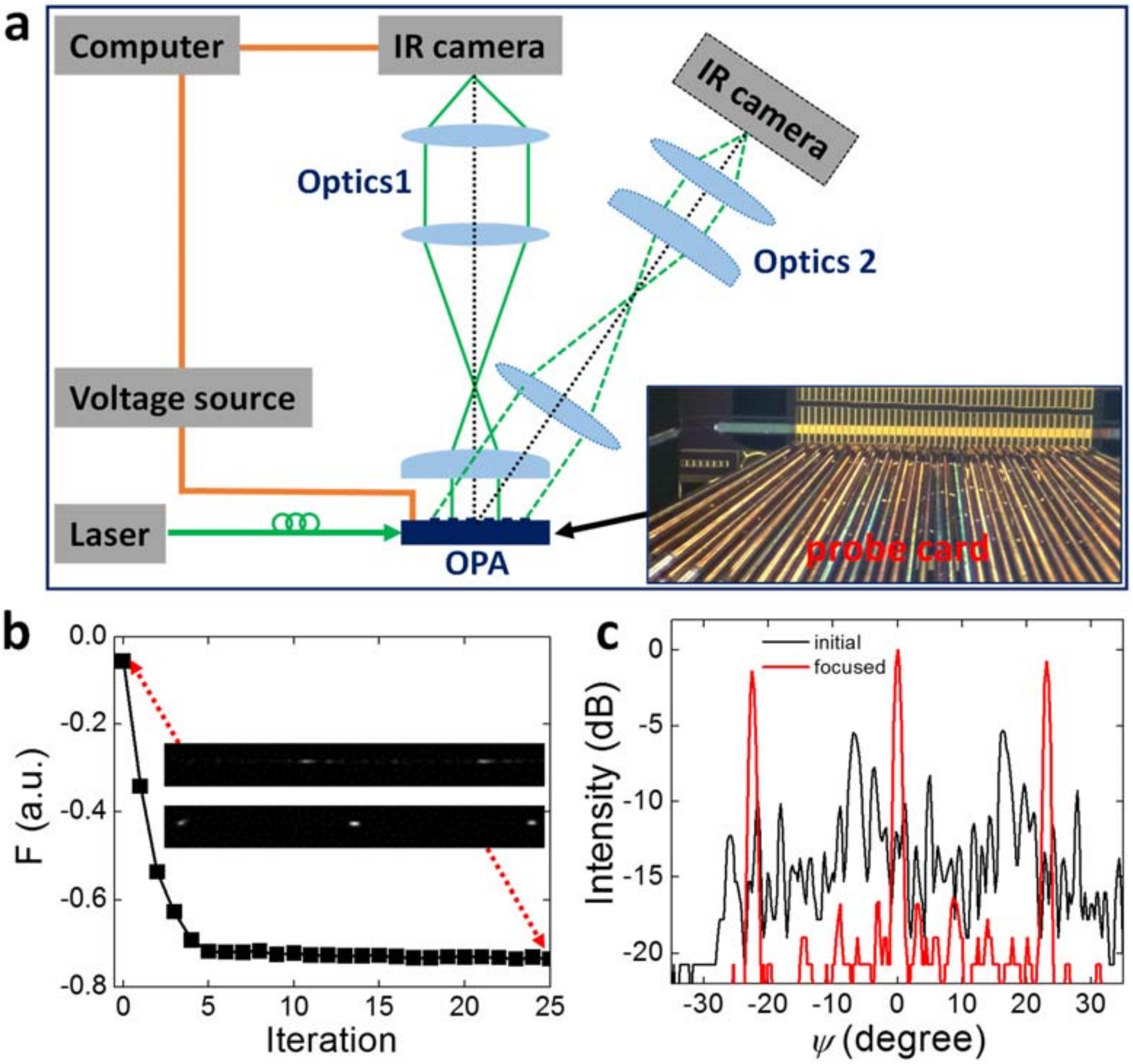

**Figure 3.** Phase alignment and beam focus. a) Fourier imaging setup for beam optimization with two optical systems. One stationary optics (Optics 1) has a fixed optical axis normal to the chip surface and a high numerical aperture (NA) of NA = 0.83 with a far field resolution of 0.3°. The other (Optics 2) is rotatable with a NA=0.23 and a theoretical resolution of 0.0043°. The rotational optics was also used to calibrate the emission angle. b) Evolution of the optimization function. The inset images correspond to the initial beam and the focused beam respectively. c) The far field beam profiles before and after phase alignment in $\psi$ axis, with a side-mode suppression ratio of 16 dB after focus.

By tuning the phases of the OPA, we can steer the beam along lateral direction in an entire steering angle range, as shown in **Figure 4**a. The full width at half maximum (FWHM) beam width was measured to be 0.78° in our 32-channel OPA and it can be easily reduced by increasing the number of elements in OPAs ($\delta\psi \approx \lambda/Np_g$, with $N$ the number of channel in the OPA). Beam steering in the longitudinal direction was realized by tuning the wavelength, according to the formula $\sin\theta = (n_{\mathrm{eff}}\Lambda - \lambda)/\Lambda$, where $n_{\mathrm{eff}}$ is the effective index of the waveguide, $\Lambda$ is the grating period. Figure 4b presents the beam profiles when the beam is steered by changing the wavelength from 1450 nm to 1650 nm with a noise floor less than -20 dB (limited by the IR camera used). The overall steering angle in $\theta$ axis is ~28° over a 200 nm wavelength, resulting in a tuning efficiency by wavelength of 0.138°/nm. The FWHM beam width was measured to be 0.02° at 1550 nm corresponding to a ~2.5 mm effective aperture size. In Figure 4c, we also plot the 2D beam profiles when steering the beam in ($\theta$, $\psi$) plane in the OPA with a 4μm-pitch grating array. Note that the beam steering angle in $\psi$ axis can be further increased by tapering down the pitch from phase-shifter array to grating array or directly reducing the OPA pitch. With the pitch-tapering scheme, we further demonstrated 51° steering range in $\psi$ axis using a 2μm-pitch grating array. In this way, however, the effective aperture size relative to the chip size in the OPA dimension is not improved. We also tried to push the phase shifter pitch down to 2 μm, but the yield for III-V diodes was low, mainly limited by lithography which could be addressed by using more advanced tools. We believe that III-V/Si OPA with pitch approaching subwavelength dimensions can eventually be pursued to realize diffraction-limited and grating-lobe-free beams. The dimension of the III-V/Si OPA can be scaled to thousands of elements by taking the advantage of CMOS scale processing.

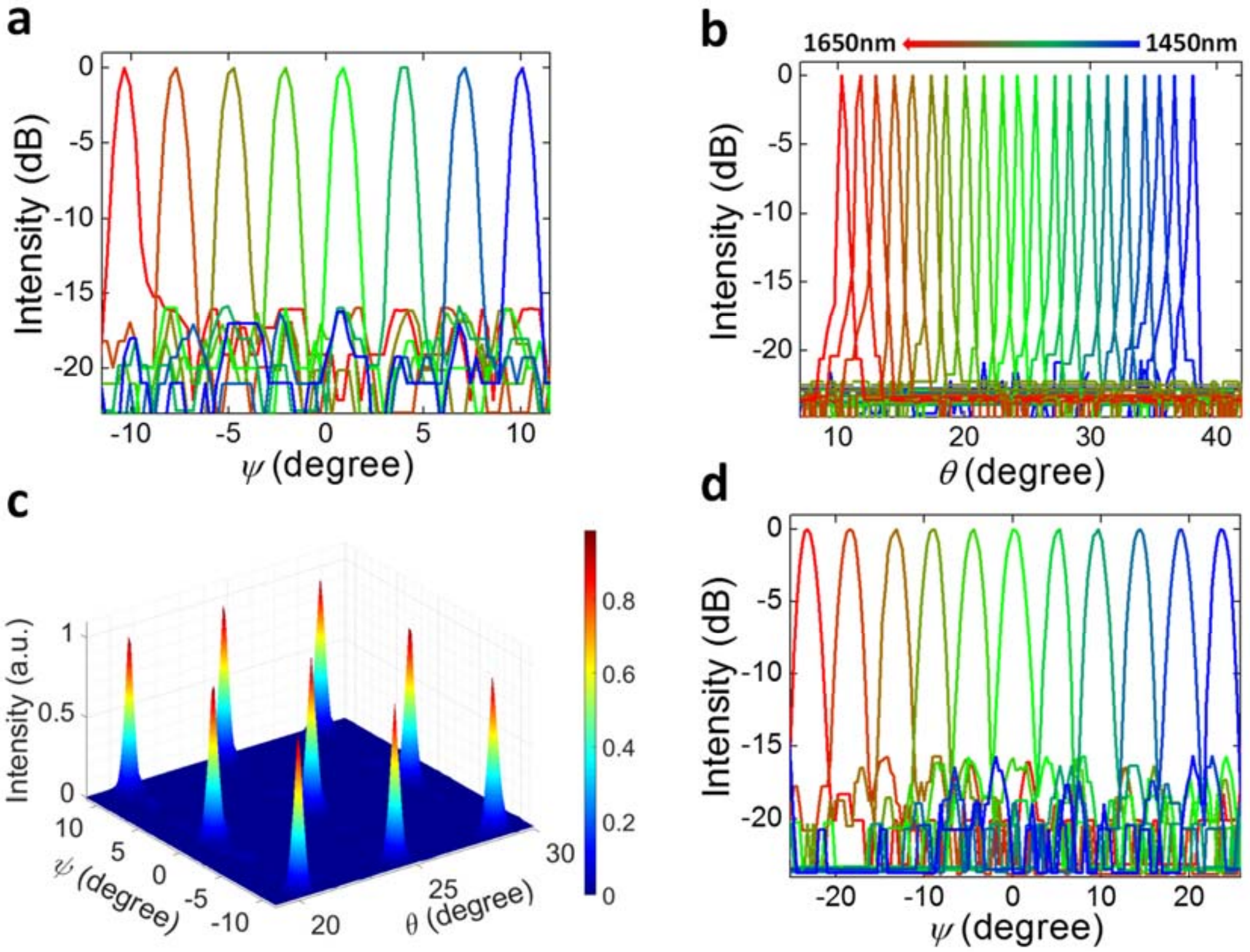

**Figure 4.** Beam steering in ($\theta$, $\psi$) plane. a) Measured beam profiles at 1550 nm wavelength with the beam steering across the field of view in the $\psi$ axis at a 2.8° increment in the OPA with a 4µm-pitch grating array. b) Beam profiles along $\theta$ axis ($\psi = 0$) with the beam steering when tuning the wavelength from 1450 nm to 1650 nm at a 10 nm increment. c) Plots of 2D beam profiles with a wavelength tuning range from 1520 nm to 1580 nm. Note that the far-field images in both $\theta$ and $\psi$ directions were acquired using low-resolution Optics 1 which limits the resolvable beam width to be ~0.3°, and additionally the plotting scale in $\psi$ axis is about two times of that in $\theta$. Thus the 3D-plotted beam appears Gaussian-like. d) Beam profiles of the OPA with a 2µm-pitch grating array and the beam steering across 51° range in the $\psi$ axis at a 4.6° increment.

## 4. Conclusion

In summary, we have demonstrated a dense III-V/Si OPA system with a pitch of 4 µm fabricated using heterogeneous wafer-bonding technology. The fabricated OPA, with III-V/Si phase shifters, offers state-of-the-art performance in terms of drive voltage for 2π phase shift (-0.9 V) and residual amplitude modulation (0.15 dB for 2π phase shift). Beam steering with a steering-angle range of 28°×51° in the 32-channel OPA with a grating-array pitch of 2 µm, and a far field beam with a side-mode suppression ratio of 16 dB were demonstrated. Finally, we point out that this is a very promising approach for realizing energy-efficient and scalable III-V/Si OPAs with lager apertures. Moreover, fully integrated OPA-based systems such as LiDARs that also include laser sources, amplifiers and detectors can be realized on the same heterogeneous integration platform.

## Supporting Information

Supporting Information is available.

## Acknowledgements

The authors thank Larry Coldren and Jonathan Klamkin of UCSB, Paul Suni and James R. Colosimo from Lockheed Martin for useful discussions, and MJ Kennedy and Alfredo Torres for process help. This work is supported by DARPA MTO (MOABB HR0011-16-C-0106). The views, opinions and/or findings expressed are those of the author and should not be interpreted as representing the official views or policies of the Department of Defense or the U.S. Government.

# Supporting Information

**Dense III-V/Si Phase-Shifter Based Optical Phased Array**

*Weiqiang Xie*, Tin Komljenovic, Jinxi Huang, Michael L. Davenport and John E. Bowers*

## S1. Design of the III-V/Si phase shifter in OPAs

On-chip OPAs can contains thousands of elements and therefore require high-performance phase shifters providing low operating voltage and low power consumption, low optical loss, and large bandwidth. In Si-based phase shifters, it remains hard to overcome all the trade-offs due to an inherent low plasma-dispersion effect and considerable carrier absorption in Si [1]. In contrast, III-V phase shifters can utilize additional modulation effects, such as quantum-confined Stark effect (QCSE), band-filling, and Pockels effects among others, which increases the total electro-optic effect [2-6], leading to a low operating voltage. In addition, the higher electron mobility in III-V [5] contributes to reduced carrier absorption loss and lower resistance–capacitance time constant, allowing for low optical loss and large bandwidth. Based on our previous findings [2-4, 7], we adopt III-V multiple quantum well (MQW) PN diodes on Si waveguides to form a heterogeneous III-V/Si phase shifter. Compared with our previous work on the III-V/Si modulator which had a single modulator device and aimed for very high-speed optical modulation (a few tens of GHz), the current work focuses on the dense integration of a large number of III-V/Si phase shifters in an OPA. In this context, there are two essential problems to be addressed – the design of high-performing phase shifter in dense configuration and the feasibility of fabrication. Moreover, the performance requirements for phase shifters in OPAs is quite different from our previous applications, for instance, low residual amplitude modulation, low power consumption and low operating voltage are more important while bandwidth requirements can be relaxed to few GHz or less, which is still sufficient for OPA applications that are usually limited by time-of-flight delay of the optical beam.

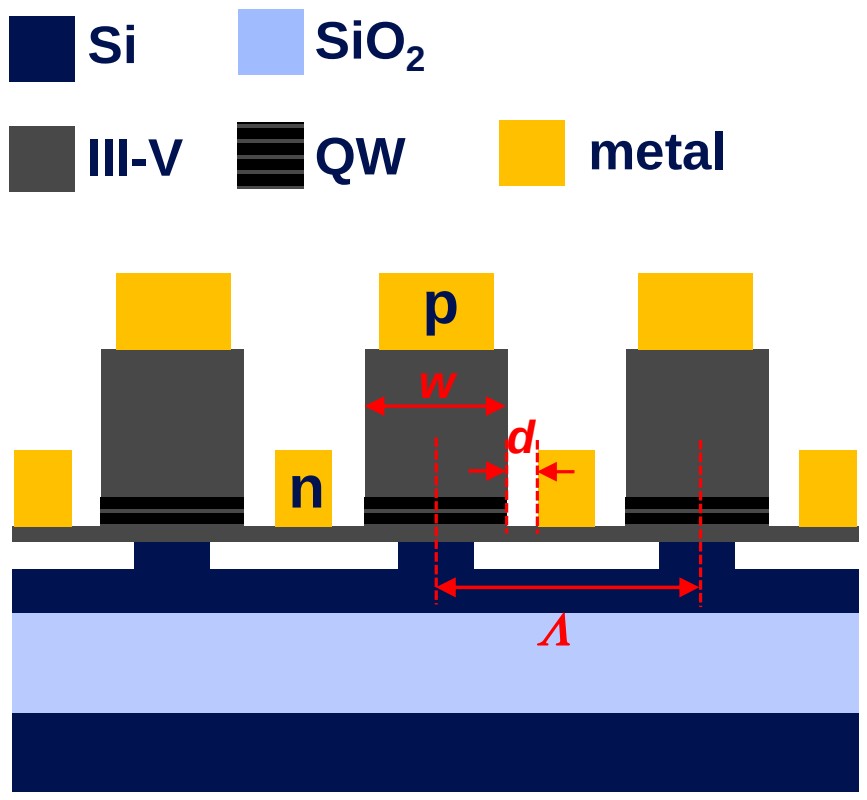

Figure S1. Schematic cross-section of the III-V/Si phase shifter array. The array pitch is $\Lambda = 4\mu m$, the III-V mesa width is $w = 4\mu m$, and the distance between the mesa boundary and n metal boundary is $d$.

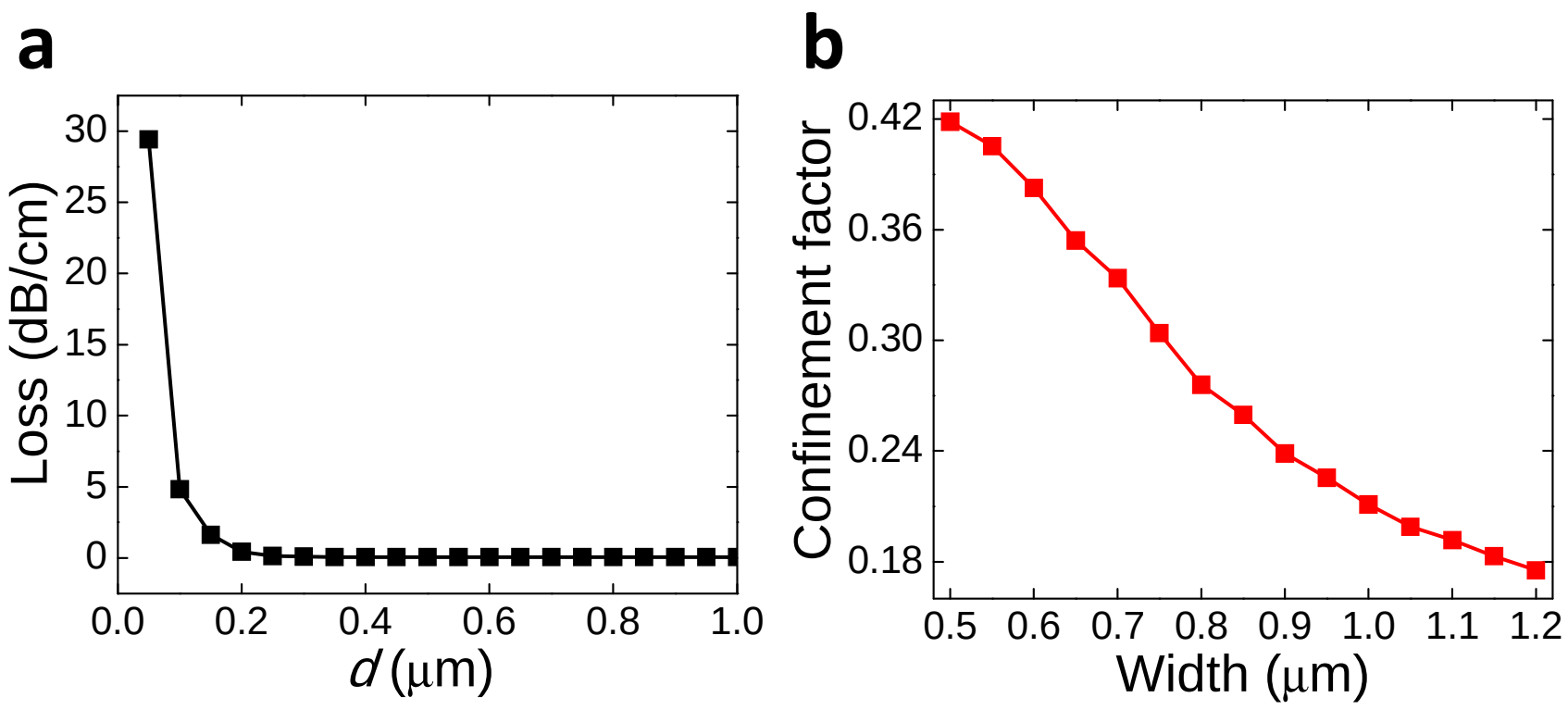

Figure S2. (a) The mode loss as a function of $d$. The Si waveguide width is set to 600 nm. (b) The confinement factor of the mode in the QW as a function of Si waveguide width.

Figure S1 shows the cross section of the design of III-V/Si phase shifter array with a pitch of 4 μm. For such dense heterogeneous integration, the room for the design is very limited and all dimensions should be considered carefully. First, we chose the III-V mesa width of 2 μm so that both the III-V process and n/p metallization can have considerable freedom for fabrication tolerance. Second, the contact metal especially for n contact needs careful design, since when the metal is close to the III-V/Si waveguide the mode loss due to adjacent metal absorption becomes significant. We calculated the mode loss as changing the distance between n contact and mesa and plot the results in Figure S2(a). It can be seen that when the distance is larger than 200 nm the loss is less than 0.5 dB/cm. In the design, a distance of 600 nm is used to relax the requirement for lithography and alignment. Another parameter to be considered is the width of Si waveguide underneath the III-V, which determines the confinement factor of the mode in QW where the modulation occurs. Figure S2(b) shows the confinement factor of the mode as a function of waveguide width. A high confinement factor is desired to maximize the modulation efficiency and thus a narrow waveguide is preferred. Since the loss of passive waveguide also increases when reducing the width, we choose an appropriate width of 600 nm in the design.

To lower the operating voltage for 2π phase shift, we choose a relatively long phase shifter with a length of 5 mm. On the mask, we also include the design for the array with a pitch of 2 μm where both the metal and III-V dimensions are scaled down by a factor of 2.

## S2. Characterization of the single waveguide grating

In our OPA demonstration, we adopt a surface grating aligned on the top of the ridge waveguide to emit light and in design the grating has a width of 800 nm and a period is 560 nm with a uniform duty cycle of 0.5. Unlike a traditional vertical grating coupler which has an effective emission length within tense of micrometers, our surface grating is shallow etched with an etching depth of ~14 nm in order to have a weak radiation strength (~28 dB/cm) and hence a long effective emission length (at least 2 mm for 14 nm etch). Note that the total length of the grating is 10 mm on the mask in order to achieve even longer effective emission length which can be achieved with further optimization of the grating design.

Since the radiation strength highly depends on the etching depth, we developed a dry etch process with very low etch rate of ~12 nm/min for Si to control the etched depth accurately, as shown in Figure S3. We characterized the far field beam of a single waveguide grating using a Fourier imaging setup. Figure S4(a) shows the CCD images of the far field beams at different wavelengths. It can be seen that the beam is narrow in one direction but divergent in the other direction due to the limited aperture in lateral dimension (~800 nm). In an OPA, the beam can be focused in this direction by aligning the phases. The emission angle of the beam can be tuned by changing the wavelength and we obtained a tuning efficiency in longitudinal emission angle of 0.138 ± 0.002°/nm, which is in excellent agreement with a theoretical value of 0.139°/nm. We also use this method to tune the steering angle of the beam in one dimension in the OPA. The FWHM beam width was measured to be 0.02° at 1550 nm by fitting a Gaussian function, corresponding to a ~2.5 mm effective emission length. A narrower beam can be achieved by optimizing the emission strength profile or using even weaker grating emitters and the relevant work can be found in our recent work [8].

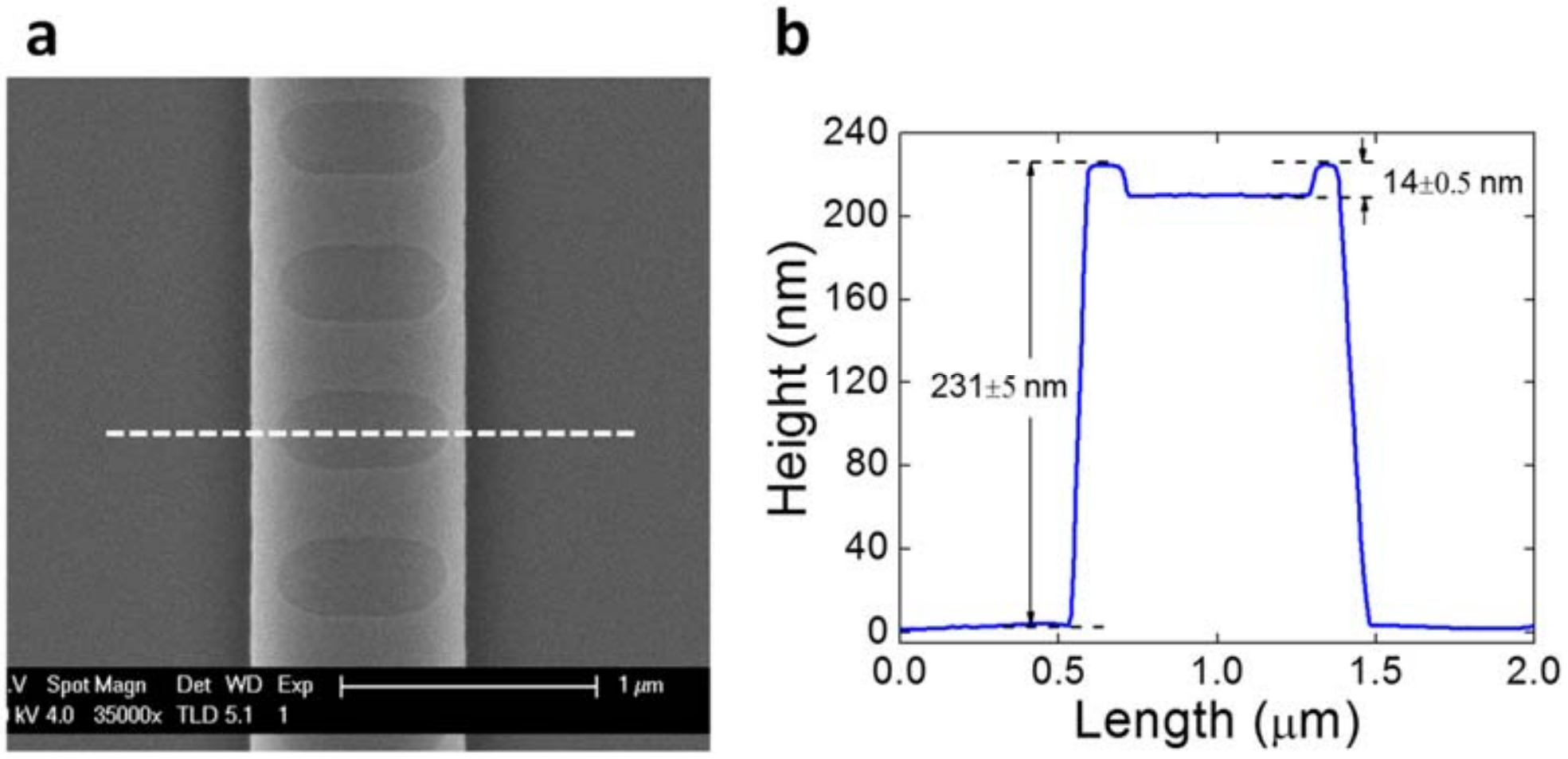

Figure S3. (a) SEM image of the fabricated waveguide grating. (b) AFM measured height profile of the grating, showing the waveguide height of 231±5 nm and the grating depth of 14±0.5 nm.

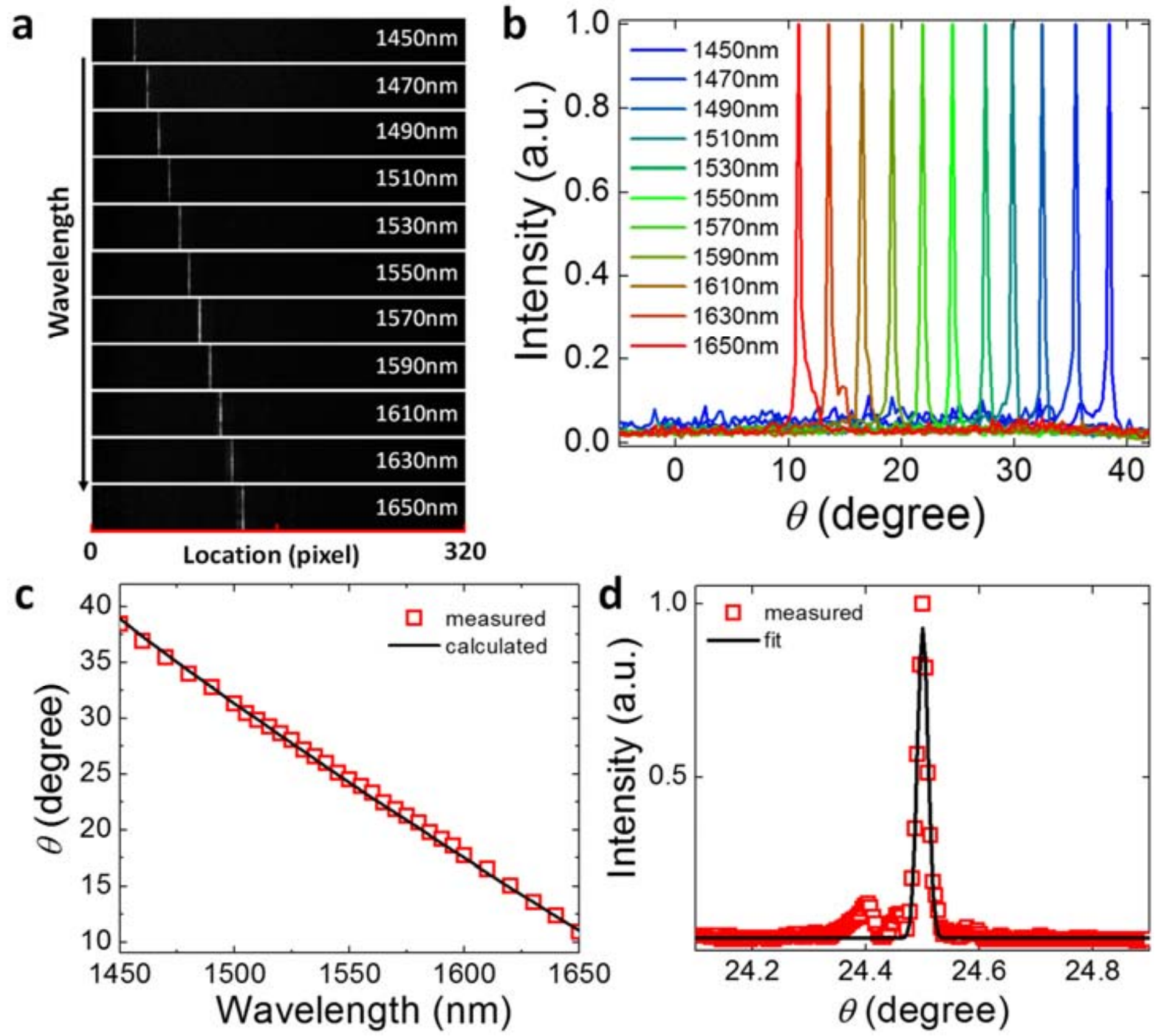

Figure S4. (a) CCD images of the far field beam. (b) Beam profiles for different wavelengths. (c) Measured and calculated emission angle $\theta$ as a function of wavelength. (d) High-resolution beam profile at the wavelength of 1550 nm. The fitted FWHM beam width is 0.02°.

## S3. Fabrication of OPAs

The devices were fabricated on a 100 mm SOI wafer using 248 nm DUV lithography and a III-V/Si heterogeneous integration technique [7]. We first patterned the Si waveguide layer and then bond the III-V die onto Si. The III-V processing includes the definition of mesa by a RIE etch, QW etch, and n-contact region etch. Finally, the n/p contact metallization and probe metal process were performed. Due to the small feature size of the dense structure and the large topology (> 2 µm) of the surface, we developed a triple-layer process for metal lift-off as shown in Figure S5. In this fabrication, first a photoresist is patterned on a polymer/$SiO_2$ bi-layer and an undercut profile is then created by dry etching the polymer with $SiO_2$ as a mask; lastly, a metal layer for contact is deposited in an electron-beam evaporation system and then the lift-off is done in a solvent. The developed process allows us to define dense metal lines with a pitch down to 2 µm or even less and a thickness of up to 1 µm on a flat substrate.

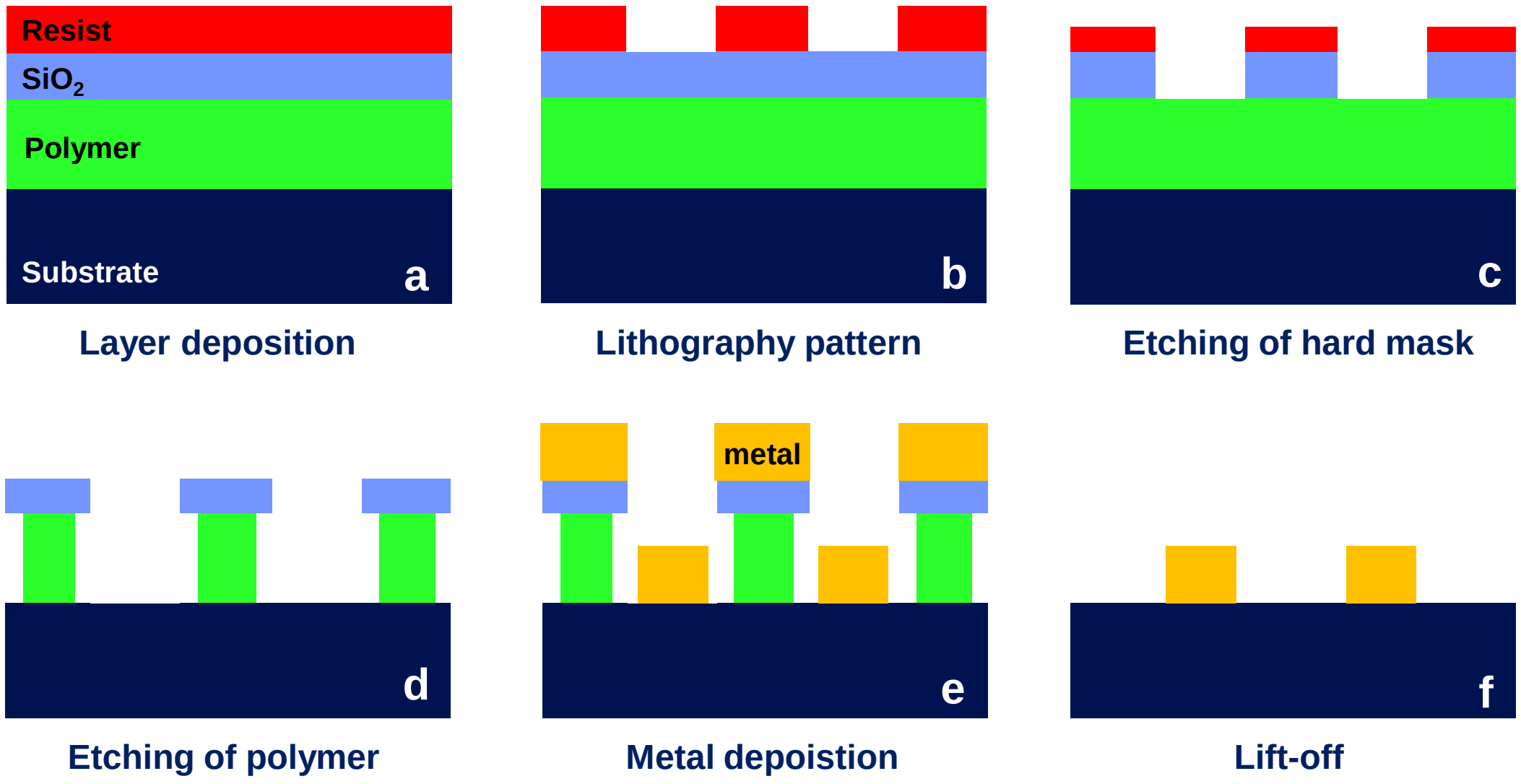

Figure S5. The flow of the triple-layer process for metal lift-off on a flat substrate.

In Figure S6, we show the photograph of one fabricated chip diced from the 100 mm wafer. In Figure S7, we show the cross sections of the fabricated phase shifter array, demonstrating high quality of the process and cross-section following the designed structure. The yield for fabricating the III-V/Si phase shifters mainly depends on the bonding quality of III-V epi on patterned Si surface. For 32-channel OPAs, the quality is high due to small area and thus this yield can be up to 100% as we achieved in experiment. For an even larger scale OPAs, the yield can be improved by optimizing bonding conditions and pre-screening the surface quality of both Si and III-V material. For smaller pitch array (2 µm), the yield was low - mainly limited by lithography, which can be improved by using more advanced tools in future.

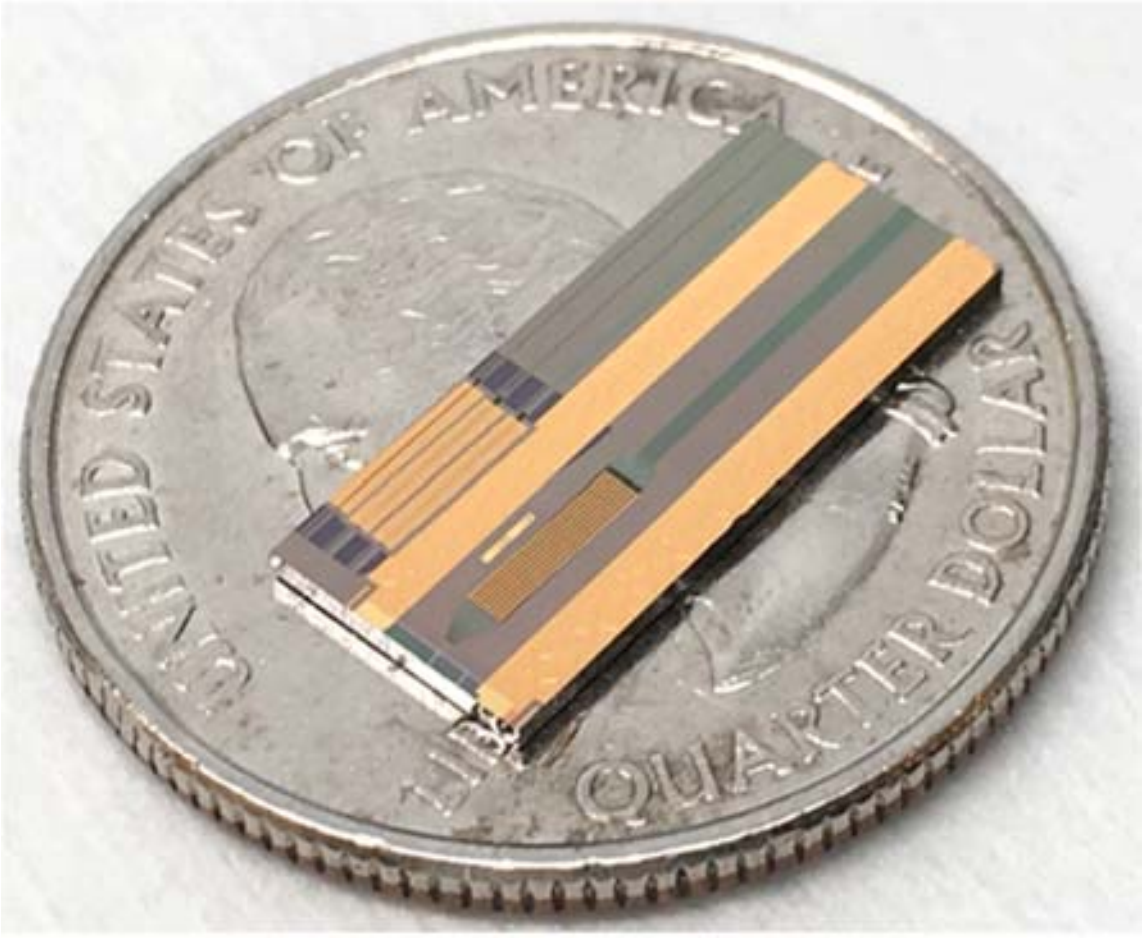

Figure S6. A set of three 32-channel and one 240-channel OPAs on the top of a US quarter coin for size comparison.

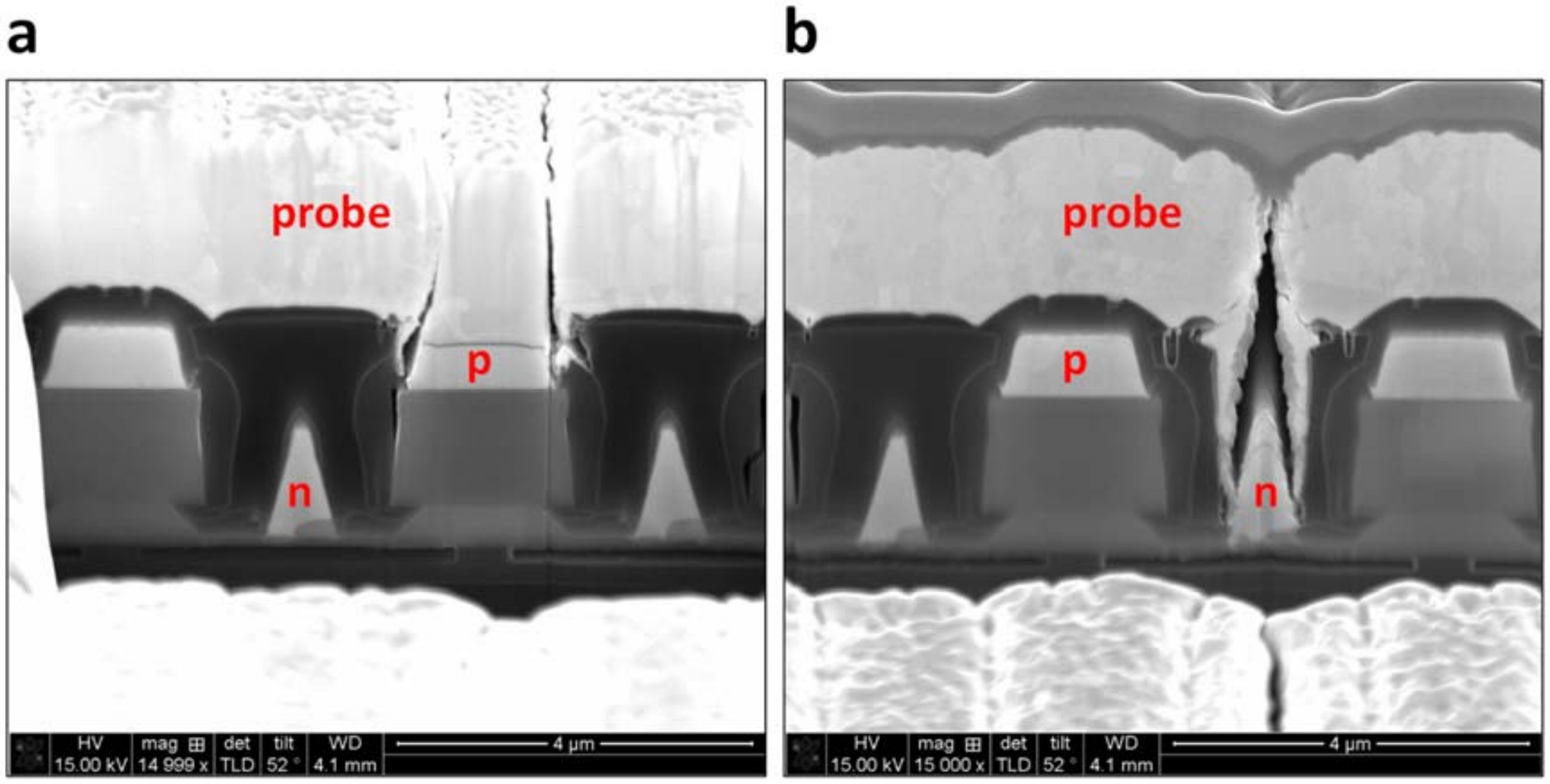

Figure S7. The cross-sectional images of the III-V/Si phase shifter array with locally opened p (a) and n (b) contact for probe metal in one shifter on top.

## S4. Characterization of the phase shifters

Figure S8(a) shows the I-V curve of III-V diode in a MZI modulator and a dark current at -2 V bias is only 3 nA, implying a power consumption at a level of nanowatts for the phase shifter. In Figure S8(b), we show the measured frequency response of the MZI modulator and obtain a 3dB electrical bandwidth of ~1.65 GHz limited by capacitance. Such bandwidth is sufficient for OPA applications. In Figure S9, we also show the I-V curves for a 32-channel. It can be seen that the electrical performance of III-V diode is uniform and the dark current is in a range of 1-3 nA at -1 V bias, which implies the power consumption on the order of nanowatts for all phase shifters. The uniformity of the performance in return proves the reliability of the heterostructure III-V/Si platform for large-scale integration.

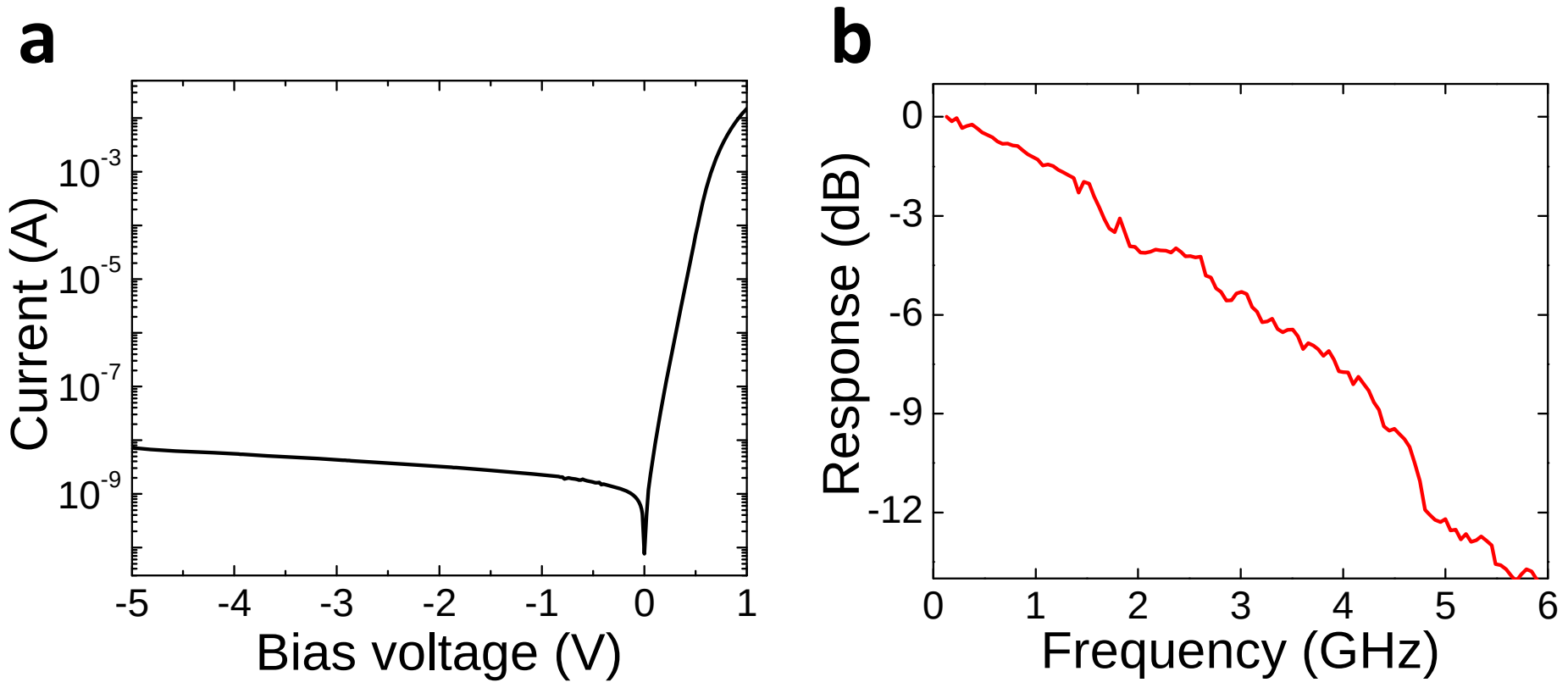

Figure S8. (a) I-V characteristics of the phase shifter. (b) Frequency response of the MZI modulator.

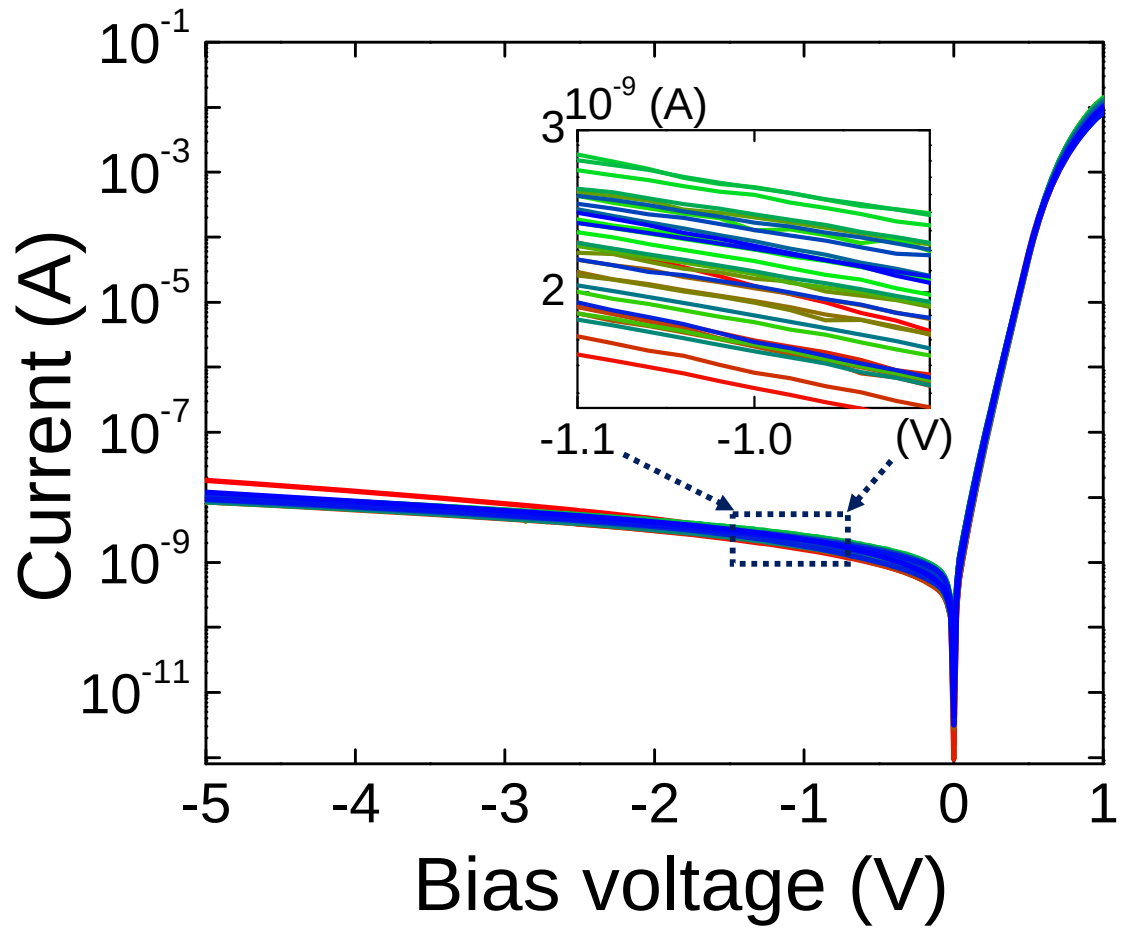

Figure S9. I-V curves of the 32-channel phase shifters. The inset shows the enlarged plot around -1 V bias.

In OPAs, a low voltage–length product ($V_\pi L$) is desired for phase shifter in order to relax the requirement of the electrical driver for the OPA. One way to reduce the $V_\pi L$ is increasing the confinement factor of the mode in III-V QW region as we discussed before. In Figure S10, we plot the measured $V_\pi L$ at different confinement factors for the waveguide widths of 1000 nm, 800 nm, and 600 nm. It is clear that a lower $V_\pi L$ can be achieved at a higher confinement factor and for 600 nm waveguide the $V_\pi L$ of 0.225 V·cm is obtained. Therefore, in our OPA with 5 mm phase shifters, the operating voltage for 2π phase shift is less than -1.0 V and the power consumption is on the order of 1-3 nanowatts for single phase shifters, which is, to the best of our knowledge, a record low operating voltage and low-power on-chip OPA.

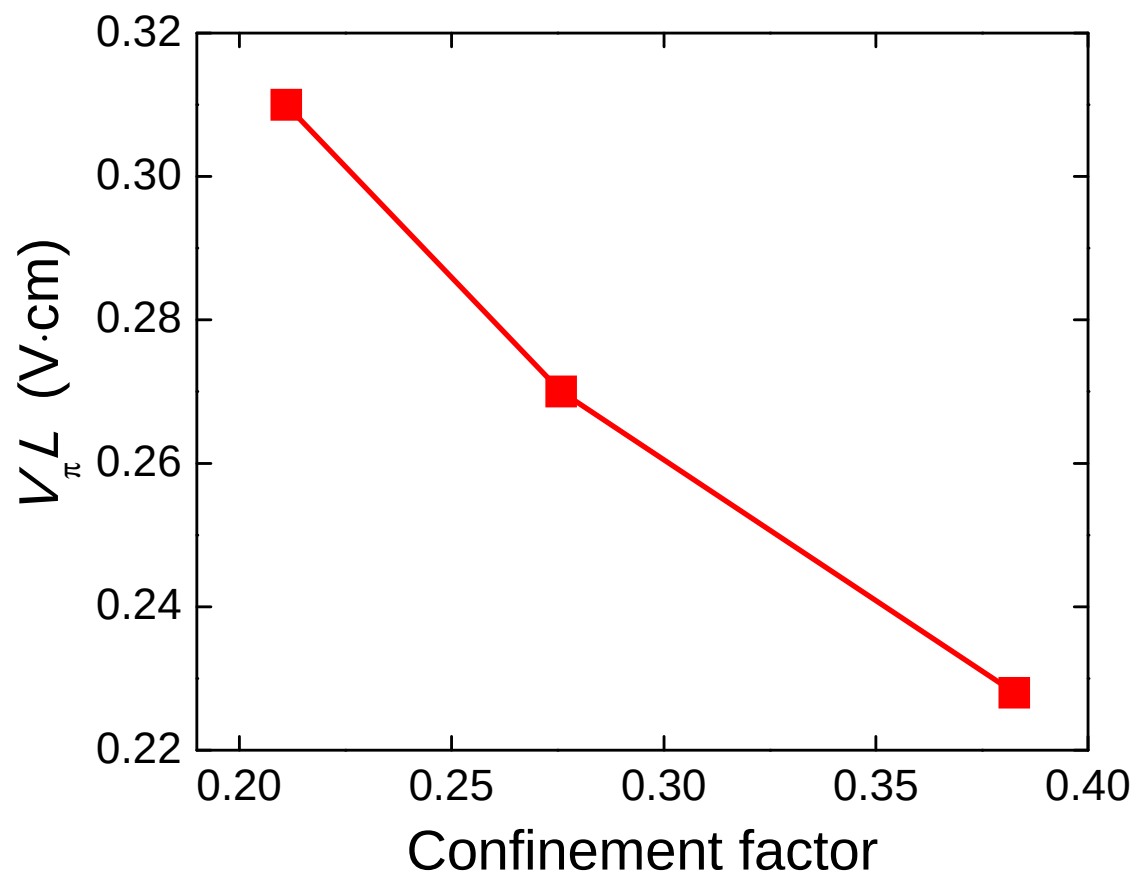

Figure S10. $V_\pi L$ as a function of confinement factor for the waveguide widths of 1000 nm, 800 nm, and 600 nm. The phase shifter was reversely biased and the data were measured at 1550 nm wavelength.

## S5. Characterization of the far-field beam width

The far-field beam width of the OPA in both $\psi$ and $\theta$ directions is wavelength-dependent. In $\psi$ axis, the beam width is determined by $\delta\psi \approx \lambda/Np_g$, with $\lambda$ the free-space wavelength, $N$ the number of channel and $p_g$ the pitch of grating array. In $\theta$ axis, the beam width is related to the effective emission length of the grating which depends on the wavelength. We show the measured FWHM beam width at different wavelengths in Figure S11. When increasing the wavelength, the beam width in $\psi$ broadens, agreeing with the theoretical prediction. In $\theta$ direction, within our measurement range the narrowest beam width was measured at 1580 nm, which means the longest effective emission length.

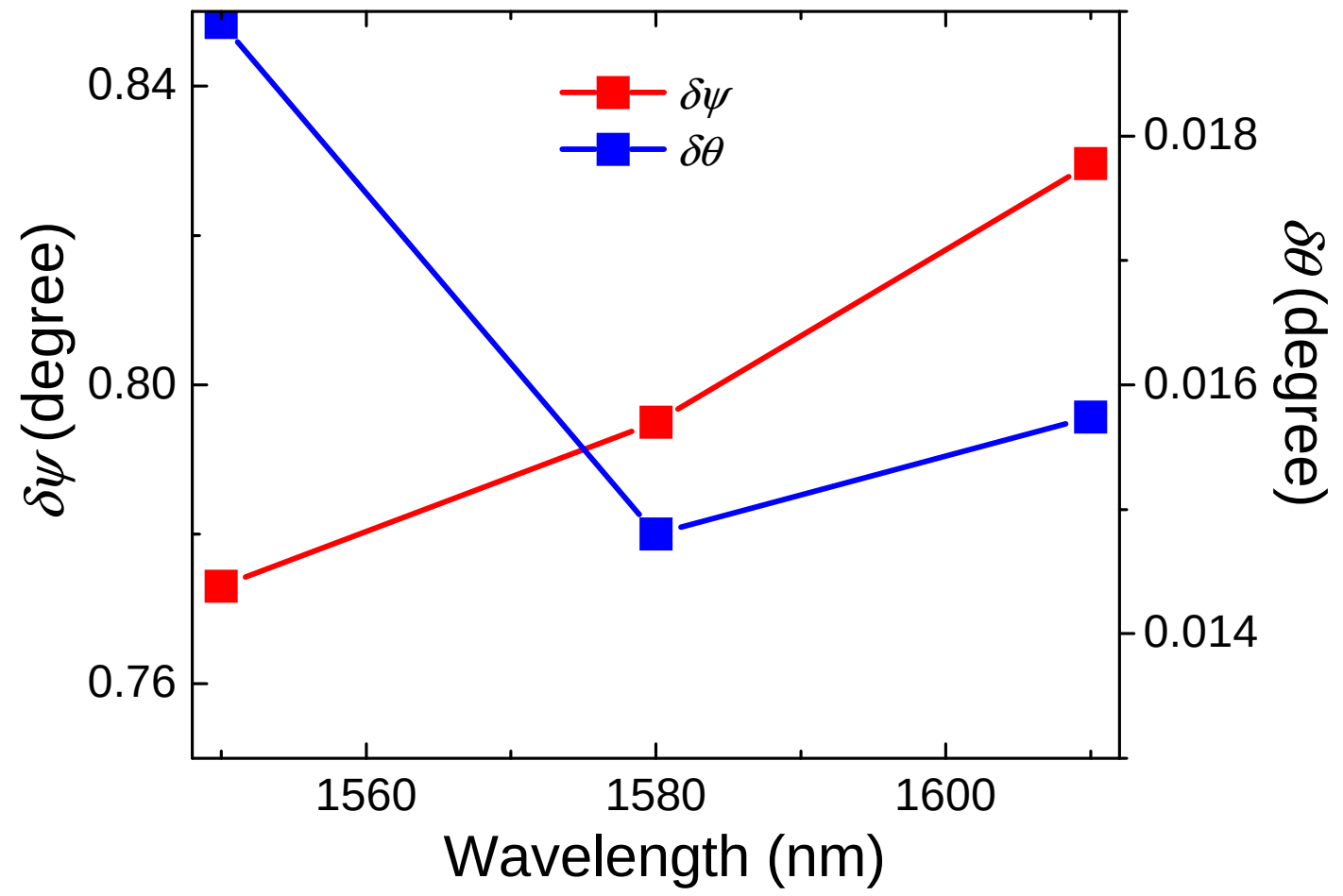

Figure S11. The measured beam width in both $\psi$ and $\theta$ directions at different wavelengths.